\documentclass[superscriptaddress,nobibnotes,amsmath,amssymb,notitlepage,twocolumn,pra,longbibliography]{revtex4-1}

\usepackage{pdfpages} % for appending SM at end of main text
\usepackage{etoolbox}

\usepackage{bm,mathptmx,braket}
\usepackage{graphicx,color,hyperref}
\usepackage[caption=false]{subfig}
\hypersetup{colorlinks=true, linkcolor=blue, citecolor=blue, urlcolor=blue} 

\graphicspath{{./img/}}

\newcommand{\beq}[1]{\begin{equation}\label{#1}}
\newcommand{\eep}{\;.\end{equation}}
\newcommand{\eec}{\;,\end{equation}}
\newcommand{\eeq}{\end{equation}}

\newcommand*\dd{\mathop{}\!\mathrm{d}} %differential d
\newcommand{\pd}[2]{\frac{\partial#1}{\partial#2}}

\newcommand{\lb}{\left(}
\newcommand{\rb}{\right)}

\newcommand*\chem[1]{\ensuremath{\mathrm{#1}}} % for chemical symbols
\renewcommand{\a}{\alpha}
\renewcommand{\b}{\beta}
\newcommand{\g}{\gamma}

\renewcommand{\k}{\kappa}
\newcommand{\la}{\lambda}
\renewcommand{\th}{\theta}

\newcommand{\om}{\omega}

\newcommand{\D}{\Delta}

\newcommand{\Om}{\Omega}

\DeclareMathAlphabet{\mathcal}{OMS}{cmsy}{m}{n} % Changes font for mathcal but leaves the rest of the math fonts in Times.

\newcommand{\Df}{\mathcal{D}}    % deformation field
\newcommand{\scBZ}{\text{scBZ}}
\newcommand{\Omsc}{\Om_{\text{sc}}}

\newcommand{\occ}{\text{occ}}

\renewcommand{\vec}[1]{{\bf #1}}

\newcommand{\x}{\vec{x}}
\newcommand{\kv}{\vec{k}}
\newcommand{\rv}{\vec{r}}

\newcommand{\bv}{\vec{b}}
\newcommand{\R}{\vec{R}}

\newcommand{\w}{\vec{w}} % Wannier center (is a vector)

\renewcommand{\P}{\vec{P}}   % Polarization, vector
\newcommand{\kperp}{k_{\perp}}

\definecolor{orange}{rgb}{1,0.5,0}

\makeatletter
\patchcmd{\@outputpage@head}{\@ifx{\LS@rot\@undefined}{}{\LS@rot}}{}{}{}
\makeatother

\begin{document}

\title{Polarization textures in crystal supercells with topological bands}

%%%% AFFILIATIONS %%%
\newcommand{\TCM}{Theory of Condensed Matter, Cavendish Laboratory, University of Cambridge, J.\,J.\,Thomson Avenue, Cambridge CB3 0HE, United Kingdom}
\newcommand{\HarvardSeas}{John A.~Paulson School of Engineering and Applied Sciences, Harvard University, Cambridge, Massachusetts 02138, USA}

%%% AUTHORS %%%

\author{Wojciech J. Jankowski}
\affiliation{\TCM}

\author{Daniel Bennett}
\affiliation{\HarvardSeas}

\author{Aneesh Agarwal}
\affiliation{\TCM}

\author{Gaurav Chaudhary}
\affiliation{\TCM}

\author{Robert-Jan Slager}
\affiliation{\TCM}

\date{\today}

\begin{abstract}

Two-dimensional materials are a highly tunable platform for studying the momentum space topology of the electronic wavefunctions and real space topology in terms of skyrmions, merons, and vortices of an order parameter. 
Such textures for electronic polarization can exist in moir\'e heterostructures. 
A quantum-mechanical definition of local polarization textures in insulating supercells was recently proposed. 
Here, we propose a definition for local polarization that is also valid for systems with topologically nontrivial bands. 
We introduce semilocal hybrid polarizations, which are valid even when the Wannier functions in a system cannot be made exponentially localized in all dimensions.
We use this definition to explicitly show that nontrivial real-space polarization textures can exist in topologically nontrivial systems with non-zero Chern number under (1) an external superlattice potential, and (2) under a stacking-induced moir\'e potential. In the latter, we find that while the magnitude of the local polarization decreases discontinuously across a topological phase transition from trivial to topologically nontrivial, the polarization does not completely vanish.
Our findings suggest that band topology and real-space polar topology may coexist in real materials.

\end{abstract}

\maketitle

\section{Introduction}
The understanding and control of exotic electronic states of matter is one of the central aims of condensed matter physics. 
One notable avenue in this regard is the study of topological materials, hosting anomalous bulk and boundary effects and protected edge currents~\cite{Rmp1,Rmp2}. 
Topological insulators and semimetals are promised to affect technological advancements, with applications ranging from spintronics to possibly providing platforms for quantum computing~\cite{armitage2018,nayak2008}.
This field was arguably initiated by the observation that even without a net magnetic field, Hall responses can be achieved in the form of  
quantum anomalous Hall effects (QAHE)~\cite{haldane1988model}. 
In such QAH systems, wavefunctions exhibit a nontrivial winding characterized by a topological invariant known as a Chern class, which is an archetypal example of a characteristic class associated with complex vector bundles. 

On a seemingly different note, there has been a lot of recent interest in engineering exotic states via stacking engineering of layered materials. 
Combining layers with relative twist angles or lattice mismatches to form superlattice structures known as moir\'e materials \cite{Bistritzer2011} can lead to interesting phenomena such as superconductivity \cite{cao2018unconventional,lu2019superconductors,saito2020independent}, 
Mott-insulating behavior \cite{cao2018correlated}, ferroelectricity \cite{li2017binary,yasuda2021stacking,bennett2022electrically,bennett2022theory,ko2023operando}, 
nontrivial topology, both of bands \cite{koshino2018maximally,po2018origin,song2019all,po2019faithful,wu2021chern} 
and real space quantities including polarization \cite{bennett2023polar,bennett2023theory}, twist fields \cite{engelke2023non} and magnetic fields \cite{guerci2022designer}.
A~favorable aspect of such stacking-engineered phases is that they can in principle be tuned through the supercell period (twist angle or lattice mismatch), number of layers, and chemistry (changing the materials).
Topological states can be engineered with constituent materials which are ordinarily trivial, such as transition-metal dichalcogenides (TMDs), e.g.~\chem{MoS_2}, \chem{WSe_2}, etc. \cite{angeli2021gamma,zhang2021spin,li2021quantum}, 
where fractional Chern states at zero magnetic field have also been predicted \cite{PhysRevLett.124.106803,ledwith2020fractional} and recently experimentally observed \cite{xie2021fractional,cai2023signatures,park2023observation}.  
Moreover, because of the additional length scale of the superlattice potential, locally nonzero Chern numbers can be found in different stacking domains within the moir\'e superlattice~~\cite{PhysRevB.98.035404, guerci2023chern,guerci2023nature,proofs1A11ref,xia2023helical}.
The idea that such a topological invariant can be attributed to regions in real space, which we refer to as ``Chern domains'', is very intriguing for applications.
For example, knowledge of such domains, and the ability to engineer domains with different Chern numbers implies that edge currents can be induced and controlled on the domain walls separating them.
The topological nature of a Chern domain is locally reflected by the presence of QAHE at the domain walls~\cite{PhysRevB.98.035404} and they can be computationally characterized by Chern markers \cite{resta1998quantum, Bianco_2011}.

In moir\'e heterostructures of nonelemental compounds, the crystalline superlattices can very naturally break the inversion symmetry $\mathcal{I}$ within a domain, offering a natural platform for the development of polarization textures, which also can support topological features therein. 
Such topological polarization textures realizing merons or skyrmions, corresponding topologically to the $\pi_2[S^2] \cong \mathbb{Z}$ homotopy, were predicted in stacked bilayers of hexagonal boron nitride (hBN) under twist or strain \cite{bennett2023polar,bennett2023theory}.
Similar topological polar textures are also commonly observed in perovskite nanostructures \cite{RevModPhys.95.025001}, and were recently also realized in perovskites layered under moir\'e geometry \cite{SanchezSantolino2024}. 
Currently, it is not clear whether topological polarization textures can coexist with momentum-space band topological features. For example, the notion of localizability breaks down in topologically nontrivial bands in two or higher dimensions, where one cannot describe the electronic states using a basis of exponentially localized Wannier functions. 
As a result, the definition of local polarization textures, as applied to a trivial insulator \cite{bennett2023theory}, is no longer applicable.

In this work, we address this problem by proposing a definition of \textit{local} electronic contribution to polarization in a Chern insulator, and showing that the real-space polar topology can coexist with band topology.
Our formulation is a natural extension to the definition of local polarization in a crystal supercell \cite{bennett2023theory}, which is not straightforward, as the Berry phase capturing the electronic contribution to the electric polarization is a global property of the system, in contrast to the ionic core contributions to the electric polarization, which take the form of local dipole moments, consistent with classical electromagnetism \cite{vanderbilt1993electric,king1993theory,resta1994macroscopic,vanderbilt2018berry}. 
We formulate the local polarization by decomposing the Berry phase in terms of semilocal hybrid polarizations (SHPs), while also making a connection to Chern topology \cite{Sinisa2009}.
In particular, we consider the evolution of the local polarization in a crystal superlattice and elucidate the correspondences between the local polarization textures, local polarization jumps~\cite{Yoshida_2023}, and the changes of the bulk state topology realized in minibands.
We show that, across a topological phase transition (TPT), the local polarization in a texture, although decreasing in magnitude, does not vanish entirely. 
\begin{figure}[t] 
\centering
\includegraphics[width=\columnwidth]{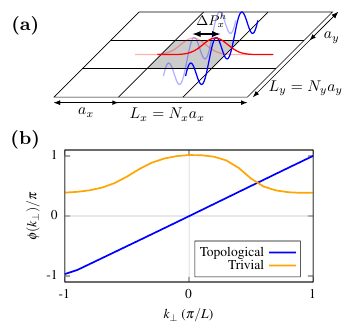}
\caption{Benchmarking the gauge-independent formulation of the local polarization in crystal supercells with topological bands.
The local polarization $P_\beta(\rv_{j})$ is defined componentwise as a sum of the semilocal hybrid polarizations (SHPs) $P^h_\beta(\rv_{j}, \kperp)$ over the perpendicular quasimomenta $\kperp$. SHPs are constituted by the shifts of the hybrid Wannier charge center (HWCCs), subject to local depolarizing perturbations $\vec{x}(\rv_{j})$. In particular, the shifts in HWCCs at every fixed $\kperp$ can be induced by a superlattice potential $V_\text{SL}$. {\bf (a)} A HWCC localized in the $x$ direction, shifting across the supercell on applying a perturbation to the central subcell, and defining $P_\beta(\rv_{j}=\vec{0})$. {\bf (b)} The Wilson loop representing the winding of Berry phase $\phi(\kperp)$ in two-dimensional supercell. The Berry phase is equivalent to the total sum of HWCCs corresponding to the occupied bands, which indicates a $total$ polarization in the supercell. On the other hand, its winding reflects the non-vanishing Chern number in the occupied topological bands. As captured by Eq.~\eqref{eq:Chern}; for supercells with topological bands there is indeed a net winding of the total hybrid polarization $P^h_\beta(\kperp)$ as a function of $\kperp$, when all occupied HWCCs are summed.
}
\label{Fig1}
\end{figure}
\section{Results}

Since our aim is to define local polarization in a periodic solid, the most natural setting is to consider a system experiencing a superlattice potential (via moir\'e engineering or external potential)~\cite{Kim2024}, such that within the supercell the polarization can acquire spatial dependence and its local definition is meaningful. 
The local polarization in a crystal supercell can be defined as the total change in the Berry phase of the supercell, subject to a local depolarizing perturbation in a given subcell starting from a non-polar reference cell configuration \cite{bennett2023theory}. 
Equivalently, the corresponding local polarization can be computed by integrating the Born effective charges along a path of phonon displacements which connect the atomic configurations in each cell.
Alternatively, the local polarization can be recast as the change in {\it all} the Wannier centers in the system with respect to the local perturbations in a given cell, as long as the Wannierized bands are topologically trivial.
For topologically trivial systems, the Wannier functions can be made exponentially localized, and the Wannier centers for each band are essentially the Berry phases, but with units of length~\cite{vanderbilt1993electric,king1993theory,marzari1997,souza2001,marzari2012maximally}. 
We briefly review these definitions of local polarization in Appendix~\ref{app::A}. It should be stressed that, contrary to the first two approaches of Ref.~\cite{bennett2023theory}, the third way via Wannier functions is not directly applicable to topological systems, which is an issue that we resolve in this work.

As mentioned above, for topologically nontrivial systems, there is an obstruction to obtaining exponentially localized Wannier functions, and the Wannier centers cannot be obtained~\cite{Sinisa2009}. 
However, we can describe the winding of the Bloch states, which is equivalent to the Berry phase, using hybrid Wannier charge centers (HWCCs) or Wilson loops (see~Fig.~\ref{Fig1}). 
The HWCCs can be obtained as the expectation values of a single component of position operator $\hat{r}_\beta$: 
\beq{}
\bar{w}^h_{n,\b}(\kperp) \equiv \bra{w^h_{n}(\kperp)} \hat{r}_\beta \ket{w^h_{n}(\kperp)}
\eec
using a basis of hybrid Wannier functions $\ket{w^h_{n}(\kperp)}$ (HWFs), which are obtained by Fourier transforming the Bloch states only in the direction $\beta$ (for more details, see Appendix~\ref{app::B}). In the case of more than two spatial dimensions, to deduce the local polarization, the Fourier transform in only one direction to obtain HWFs is similarly required.  
Here, $\kperp$ are the wavenumbers in the direction orthogonal to $\beta$. 
The total hybrid polarization in a supercell can be defined in terms of HWCCs summed over the occupied band indices $n$, 
\beq{}
P^h_{\b}(\kperp) = -\frac{ef}{\Om_\text{sc}} \sum_n^{\text{occ}} \bar{w}^h_{n,\beta}(\kperp)
\eec
where $f$ is an occupation factor, $\Om_\text{sc}$ is the supercell volume, and the HWCCs are localized in the direction $\b$. Analogously, for the purposes of defining the local polarization in a topological, non-Wannierizable crystal supercell, it is useful to introduce 
\beq{eq:P-hybrid-local}
    P^h_{\b}(\rv_j, \kperp) = -\frac{ef}{\Om_0}\sum_{n}^{\occ} \int_0^{\x(\rv_j)} \partial_{x'_{\k,\a}}\bar{w}^h_{n,\b}(\kperp)\dd x'_{\k,\a}
\eec
which we define as the semilocal hybrid polarization (SHP), where $\Om_0$ is a subcell volume (see Appendix~\ref{app::B} for more details). 
In the spirit of Ref.~\cite{bennett2023theory}, here, the integral represents the change of hybrid polarizations on introducing local displacements/reparametrizations: $\x(\rv_j) = \{x_{\k, \a}\}$, where $\k$ labels the atoms and $\a$ specifies the perturbation direction. Additionally, the Einstein summation convention was assumed. In order to deduce $P^h_{\b}(\rv_j, \kperp)$, the local perturbations are imposed only in a subcell $\rv_j$ of a supercell, bringing its configuration to the nonpolar reference state $\x(\rv_j) = \vec{0}$. The intuitive meaning of SHPs is the change of the position of the hybrid charge centers in direction $\beta$, as captured by the flow of the previously-defined polarization currents across the supercell, as the local perturbations $\x(\rv_j)$ are introduced. We further propose that the introduction of the semilocal hybrid polarizations allows us to evaluate the local polarization in a topological supercell as
\beq{eq:P-local}
    P_{\b}(\rv_j) = \oint_{\mathcal{C}} P^h_\beta(\rv_j, \kperp)\dd\kperp
\eec
where $\mathcal{C}$ is a closed loop in the BZ, starting from $\kperp^0$ and of length of the superlattice reciprocal vector $\bv_{\perp}$. 
By substituting Eq.~\eqref{eq:P-hybrid-local} into Eq.~\eqref{eq:P-local}, we obtain
\beq{P-local}
    P_{\b}(\rv_j) =  -\frac{ef}{\Om_0}\sum_{n}^{\occ} \oint_{\mathcal{C}} \int_0^{\x(\rv_j)} \partial_{x'_{\k,\a}}\bar{w}^h_{n,\b}(\kperp)\dd x'_{\k,\a} \dd\kperp
\eec
which is a natural extension to the method of computing the local polarization in terms of the Wannier functions. However, contrary to the previous definition~\cite{bennett2023theory} (see also Appendix~\ref{app::A}), Eq.~$\eqref{P-local}$ is valid for both topological and trivial bands. 
The above relation states that local polarization in a system with topologically nontrivial bands can be obtained componentwise, i.e., a certain polarization component is simply the projection of the flow of hybrid Wannier center that is exponentially localized along the same direction. 
This definition is motivated by the fact that change in polarization is the physical quantity that is fundamentally related to the polarization currents $\textbf{j}_\text{P} \equiv \dd \textbf{P} / \dd t$ flowing through the system as it is adiabatically evolved from an initial to the final state ($\Delta \textbf{P} \equiv \int \textbf{j}_\text{P} \dd \text{t} = \textbf{P}_f - \textbf{P}_i$) \cite{resta1992}. Importantly, that relation shows that electronic contribution to the electric polarization is quantum-mechanically defined as a change of net polarization induced by the flow of polarization currents, subject to an adiabatic variation of the polarization-controlling parameter, as introduced in Ref.~\cite{resta1992}.
This relation can also be resolved componentwise, allowing us to construct hybrid Wannier functions that are maximally localized in only one direction and observing their flow as the polarization currents evolve (see Appendix~\ref{app::B} for more details). The formulation of local polarization central to this work aims to reflect this intuitive picture in a maximally-local manner, as the evolving exponentially-localized HWCCs constituting the SHPs reflect the local charge flow, on having introduced the polarization-inducing local perturbations in the considered supercells.

We note that for the two-dimensional case of e.g. Chern insulators, $\beta = x, y$ specifies the in-plane directions. 
Here, $k^0_{\perp}$ should be chosen consistently for finding polarization changes, e.g.~$P^h_\b (\rv_j, \kperp^0) = 0$, when $\x(\rv_j)=\vec{0}$, which, upon choosing a maximally smooth gauge, should ensure a vanishing polarization for nonpolar configurations~\cite{bennett2023theory}. Importantly, the point $(k^0_{x}, k^0_{y})$ needs to be chosen consistently for the evaluation of $\textbf{P}(\rv_j) = (P_{x}(\rv_j), P_{y}(\rv_j))$, with the real-space integration limits of Eq.~\eqref{P-local}, which define initial and final states with respect to which the local polarization is computed as a change ($\textbf{P} \equiv \Delta \textbf{P} = \textbf{P}_f - \textbf{P}_i$)~\cite{resta1992}. If the $k$-space integral is performed inconsistently in the initial and final real-space states, the resulting polarizations acquire an erroneous term depending on the shift in the integration endpoints and the Chern number $C$, as was pointed out for arbitrary Chern insulators in Ref.~\cite{Sinisa2009}. %Importantly, the definition of the local polarization as a change of the polarization, which is due to the polarization currents $\textbf{j}_\text{P} \equiv \dd \textbf{P} / \dd t$ flowing throughout the evolution of the insulating system, from the initial to the final state ($\Delta \textbf{P} \equiv \int \textbf{j}_\text{P} \dd \text{t} = \textbf{P}_f - \textbf{P}_i$), 
The above definition is analogous to the Berry-phase formulation of the \textit{total} polarization in Chern insulators~\cite{Sinisa2009} as a global quantity. 
Indeed, upon relating HWCCs to Berry phases 
\beq{eq:Berry}
\phi_n(k_\g) = i \oint_{\kperp = k_\gamma} \braket{u_{n,\kv}|\partial_{k_\beta} u_{n,\kv}} \dd k_\beta
\eec
our definition is consistent with the previous formulations of electric polarization in Chern insulators \cite{Sinisa2009} that obtains the $total$ polarization of a topological system, without partitioning into any local contributions to the net electric dipole moment present in a supercell.

Furthermore, we can relate the SHPs to band topology in crystal supercells, therefore settling whether any information about the topological character of the minibands can be inferred from $P^h_\b(\rv_j, \kperp)$. 
It is known that the Chern number $C$ of a system can be calculated from the winding of HWCCs \cite{gresch2017z2pack}, or equivalently, hybrid polarization, along the quasimomentum component (here $k_\gamma$). 
Essentially, it is the winding of Berry phases $\phi_n(k_\gamma)$ across a Wilson loop,
\begin{align}\label{eq:Chern}
    & C = \sum_{n}^{\occ} \frac{1}{L_\b} \Big[ \bar{w}^h_{n,\b}(\kperp=2\pi) - \bar{w}^h_{n,\b}(\kperp=0)\Big]  \notag\\ 
    & \hspace{0.25cm} = -\frac{1}{efL_\b} \Big[ P^h_\b(\kperp=2\pi) - P^h_\b(\kperp=0) \Big],    
\end{align}
where we impose $L_\perp = 1$ for simplicity. 
We propose a further variant of this correspondence for the {\it local} perturbations in a~supercell, namely, 
\begin{align}\label{eq:C-loc}
    & \Delta C(\{\rv_j\}) = -\frac{1}{efL_\b} \sum_{j \in \{\rv_j\}} \left[ P^h_\beta(\rv_j, \kperp=2\pi) - P^h_\b(\rv_j, \kperp = 0) \right] \notag\\
    & \hspace{1.4cm} = -\frac{1}{efL_\b} \sum_{j \in \{\rv_j\}} \left. P^h_\beta(\rv_j, \kperp) \right|_{\kperp^0}^{\kperp^0 + \bv_{\perp} }.    
\end{align}
The natural interpretation of $\Delta C(\{\rv_j\})$ is the change of the total Chern number in the supercell minibands, as induced by a~depolarizing perturbation imposed in the chosen cells $\{\rv_j\}$. In particular, to induce a nontrivial change $\Delta C(\{\rv_j\})$, the gap between minibands close to the Fermi level must be very small, in order to admit topological phase transitions (TPTs) that cross an intermediate metallic state. Realizations of such band gaps can be naturally achieved by applying a superlattice potential to a Chern insulator, bringing it close to the critical point associated with a TPT. Under such circumstances, local displacements induced by finite-size probes, or an additional local depolarizing potential, could change the Chern number of the supercell ground state; see Fig.~\ref{Fig2}(d) for reference.

Due to the local nature of this close-to-critical setup, it is also interesting to compare it with the local Chern markers $C(\rv_j)$~\cite{Bianco_2011}, which are a real space decomposition of the total Chern number present in the minibands, reflecting the local anomalous Hall conductivity (see Appendix~\ref{app::C} for a review). 
Contrary to the conventional Chern marker $C(\rv_j)$, $\Delta C(\{\rv_j\})$ is \textit{necessarily} quantized, although restricted to supercells, i.e.~it cannot be efficiently applied to amorphous, or arbitrarily disordered systems, unless a finite size of systems supercell is assumed. The reason for quantization is that $\kperp = 0$ and $\kperp = 2\pi$ physically correspond to the same value over a compact BZ, thus also identifying the corresponding states. Hence, the HWCC flows captured by $P^h_\beta(\rv_j, \kperp=2\pi)$ need to be integer in $L_\beta$, or otherwise $\kperp = 0$ and $\kperp = 2\pi$ would be physically distinguishable. It should be noted that the regions for both quantities in  Figs.~\ref{Fig2}(d) and \ref{Fig2}(e) do not exactly coincide, despite both of the quantities being related to the band
topology of the system. The reason for that difference is that the Chern marker is a local indicator of the topology of the system, and is not quantized locally, unlike $\Delta C(\{\textbf{r}_j\})$, which is necessarily quantized. $\Delta C(\{\textbf{r}_j\})$  indicates the quantized changes in the Chern number of the supercell, reflecting the changes in the band topology, subject to an addition of depolarizing perturbations inducing TPTs.

\section{Model realizations}
\begin{figure}[t] 
\centering
\includegraphics[width=\columnwidth]{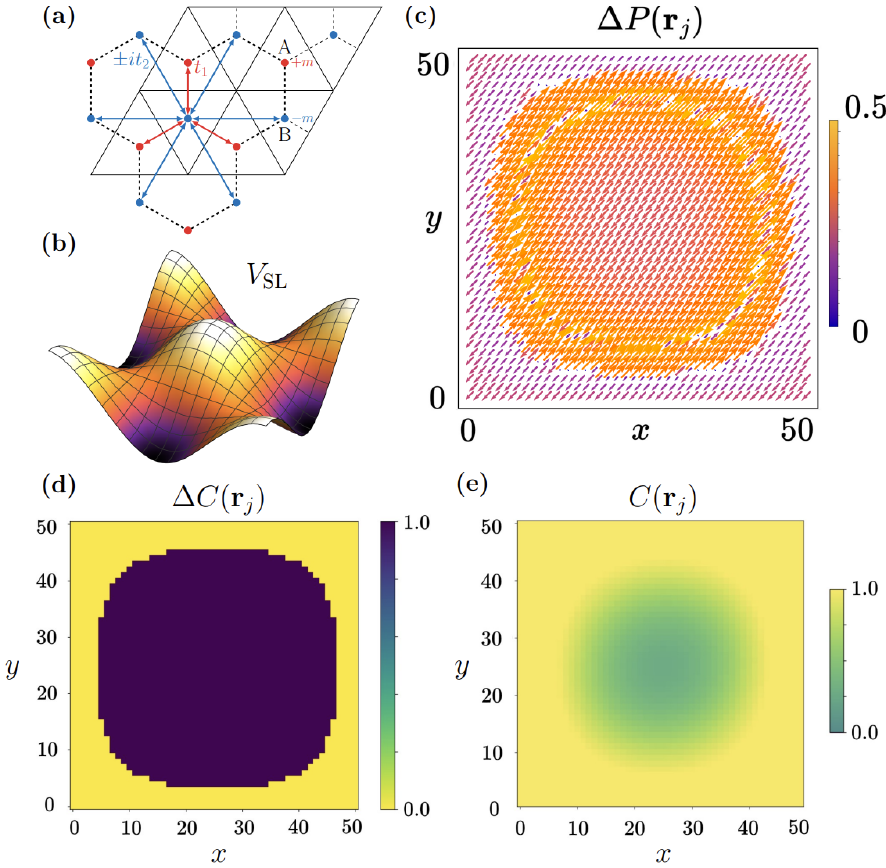}
\caption{ 
{\bf (a)} Sketch of the Haldane model. A bipartite lattice with nearest neighbor hopping $t_1$ and second nearest neighbor hopping $\pm i t_2$.
{\bf (b)} Illustration of the superlattice potential $V_{\rm SL}$.
{\bf (c)} Local polarization field in the $N=51$ super-Haldane model calculated using Eq.~\eqref{P-local}. The local polarization is defined as a change determined by the integration limits of SHPs, and is expressed in the units of the quantum of polarization~\cite{bennett2023theory}.
{\bf (d)} Quantized $\D C(\{\rv_j\})$ numbers for the $N=51$ super-Haldane model, calculated using Eq.~\eqref{eq:C-loc} for depolarizing perturbations of size $11 \times 11$ cells around each cell $j$. 
{\bf (e)} Conventional Chern markers $C({\bf r}_j)$ for the $N=51$ super-Haldane model.
}
\label{Fig2}
\end{figure}

We utilize the above theory and illustrate our findings using two examples: (i) a Chernful supercell in the presence of a superlattice potential, and (ii) a twisted moir\'e system with Chern topology, realizing a supercell with spatially modulated interlayer tunneling. 
In the first case, we consider a simple Chern insulator, namely the Haldane model [see Fig.~\ref{Fig2}(a)], subject to an addition of a superlattice potential with magnitude $V_{\rm SL}$ [see Fig.~\ref{Fig2}(b)], which we refer to as the super-Haldane model.
We first consider a honeycomb lattice with nearest ($t_1$) and second-nearest ($\pm i t_2$) neighbor hoppings, and onsite mass $\pm m$ on a~bipartite lattice of atoms $A, B$.
Further to this, within the orbital basis $A,B$, we impose the superlattice potential ${V_{Aj} = -V_{Bj} = V_{\rm SL} \cos\Big(\frac{\pi x}{N}\Big) \cos\Big(\frac{\pi y}{N}\Big)}$, where the unit cell $j$ resides at fractional coordinates $0 < x, y < N$.
The~Hamiltonian for the super-Haldane model is given by
\begin{align}\label{superHaldane}
    & H = -t_1 \sum_{\langle i,i' \rangle}(c^\dagger_{A i} c_{B i'} + \textnormal{H.c.})
    - i|t_2| \sum_{\langle\langle i,i' \rangle\rangle}(c^\dagger_{Ai} c_{Ai'} + c^\dagger_{Bi} c_{Bi'} + \textnormal{H.c.}) \notag\\ 
    & \hspace{1cm} + \sum_{j} [(V_{Bj} - m) c^\dagger_{Bj} c_{Bj} + (V_{Aj} + m) c^\dagger_{Aj}c_{Aj}],
\end{align}
where $c^\dagger_{A/B j}$ and $c_{A/B j}$ are the creation and annihilation operators for an electron in orbital $A/B$ in the cell located at $\rv_j$. Here,
$\langle ... \rangle$ denotes first neighbors, and $\langle \langle ... \rangle \rangle$ denotes second neighbors, see also Fig.~\ref{Fig2}(a). For the charge neutrality of the system to be ensured, we consider each sublattice \textit{A} and \textit{B} contributing one electron and hosting a core with a static ionic charge $Z = +1$. While we focus on the electronic contribution to the electric polarization, it is important to stress that the total gauge-invariant electric polarization of any physical system is constituted by the sum of the contributions due to electrons and ionic cores~\cite{vanderbilt2018berry}. In the considered lattice models based on the Haldane model, we assume a background of positively-charged cores at all sublattice sites, with the unbalanced onsite energies and hoppings introducing a net electric polarization through the electronic contribution. Additionally, we note that in the context of the lattice models, the change of (hybrid) polarizations, as given by Eqs.~\eqref{eq:P-hybrid-local} and \eqref{P-local}, is evaluated on introducing local reparametrizations: $\x(\rv_j)$, in the place of local displacements. Here, the reference configuration amounts to setting the onsite energy at sublattices of given cell $\rv_j$ to vanish, for more details see Appendix~\ref{app::A}.

We find that the model realizes a polarization texture [Fig.~\ref{Fig2}(c)], and there is a sharp change of the local polarization across the boundary, where the superlattice potential combined with onsite mass $m$ approaches the value of the topological mass imposed with $t_2$. The texture in Fig.~\ref{Fig2}(c) was obtained using Eq.~\eqref{P-local} for the parametrization $(t_1, t_2, m, V_{SL}) =  (1,1,0.5,10)$ of the model introduced in Eq.~\eqref{superHaldane}, with the supercell size, $N=51$. Notably, the texture demonstrates that the local polarization discontinuously flips on moving away from the supercell center, as the Haldane mass $t_2$ dominates $m$ combined with the modulated $V_{\text{SL}}$. Hence, the local polarization forms a circular domain surrounded by a visible ring, consistently with the discontinuities found between Chern and trivial insulators realized without supercells~\cite{Yoshida_2023,PhysRevLett.132.116602}.
Furthermore, we find that when superlattice potential dominates the hopping locally---effectively as a local onsite, or Semenoff~\cite{PhysRevLett.53.2449}, mass term---a reduction of the local Hall conductivity occurs. 
We support this finding by calculating the local Chern marker $C(\rv_j)$ (Fig.~\ref{Fig2}(d)), which we contrast with the quantized $\Delta C(\{\rv_j\})$ (Fig.~\ref{Fig2}(e)) introduced in the previous section. 
The changes in local Chern markers are in close correspondence with the trivialization of the SHP indicated by $\Delta C(\{\rv_j\})$, corresponding to topological transitions in response to depolarizing perturbations. 

Furthermore, we examine the connection between SHPs and band topology in another example, by stacking two monolayer copies of the Haldane model and introducing a relative twist between the layers. We refer to this system as twisted bilayer Haldanium~\cite{resta2022notes}, see Appendix~\ref{app::D} for more details.
The effective tight-binding model adapted for the studied twisted Haldanium bilayer can be compactly written as
\begin{align}\label{eq::TBtwistedH}
    & H_{\rm tHB} = [m-t_2 \sum^3_{i=1} \sin (\vec{k} \cdot \vec{b_i})] \tau_z \otimes \mathsf{1}_2 + [t_{\kv} \tau_+ \otimes \mathsf{1}_2 + \text{H.c.}]  \notag\\ 
    & \hspace{3cm} + [T(\kv, \vec{x}) \otimes \sigma_+ + \text{H.c.}].
\end{align}
Here, $\tau_i$ are Pauli matrices acting in the single-layer orbital basis ($A$,$B$), whereas the Pauli matrices $\sigma_i$ act in the top/bottom layer basis ($l=t,b$), with $\tau_+ = \frac{1}{2}(\tau_x + i\tau_y)$, and analogously for $\sigma_+$. Additionally, $\otimes$ denotes a Kronecker product, and $\vec{b_i}$ correspond to the second-neighbor hopping vectors. Importantly, $t_\kv$ represents the nearest-neighbor intra-layer hopping, while $T(\kv, \vec{x})$ is a $2~\times~2$ local stacking/configuration-dependent interlayer hopping matrix representing the tunneling of electrons between the layers, as expressed explicitly in Appendix~\ref{app::D}. Finally, in addition to the adapted tight-binding model, for more general possible studies of low-energy physics associated with topological fermions on a bilayer consisting of a honeycomb lattice, an effective continuum model of twisted Haldanium can be formulated. The model reads
\beq{eq:Ham_moire_continuum}
\begin{split}
    H_{\text{moir\'e}} = \sum_{l= \text{t,b}}\int \psi^{\dagger}_l \Bigg[\biggl(m + b \Big(\partial_{r_{\beta}}+\frac{\partial \Df_{l,\gamma}}{\partial r_\beta}\partial_{r_{\gamma}}\Big)^2 \biggl) \tau_3 \\- iv \lb \tau^{\beta} + \frac{\partial \Df_{l,\beta}}{\partial r_\gamma}\tau^{\gamma} \rb \partial_{r_{\beta}}  + v (\vec{K}\cdot \partial_{r_\beta} \Df_{l}) \tau^{\beta}\Bigg] \psi_l \dd^2\rv \\+ \int  \psi^{\dagger}_t T(\Df_{\text{t}} - \Df_{\text{b}}) \psi_b \dd^2\rv \ + \ \text{H.c.} 
\end{split}
\eec
per valley; here, without loss of generality, we consider the \textbf{K} valley (see also Appendix~\ref{app::D}). Consistently with Refs.~\cite{Balents2019,bennett2023theory}, we introduce $\Df_{\text{l}}$ as a deformation field in layer $l$, $\psi^{\dagger}_l(\rv)$, $\psi_l(\rv)$ are the fermion creation/annihilation operators, $m$/$b$ are the trivial and topological masses, while $T(\Df_{\text{t}} - \Df_{\text{b}})$ represents the interlayer tunneling. The model holds beyond the configuration space approximation \cite{carr2018relaxation,bennett2023polar}, hence as with the super-Haldane model, the polarization texture can be obtained by generalizing Eq.~\eqref{eq::TBtwistedH} to the continuum model, see Appendix~\ref{app::D} for details.
\begin{figure*}[t] 
\centering
\includegraphics[width=\linewidth]{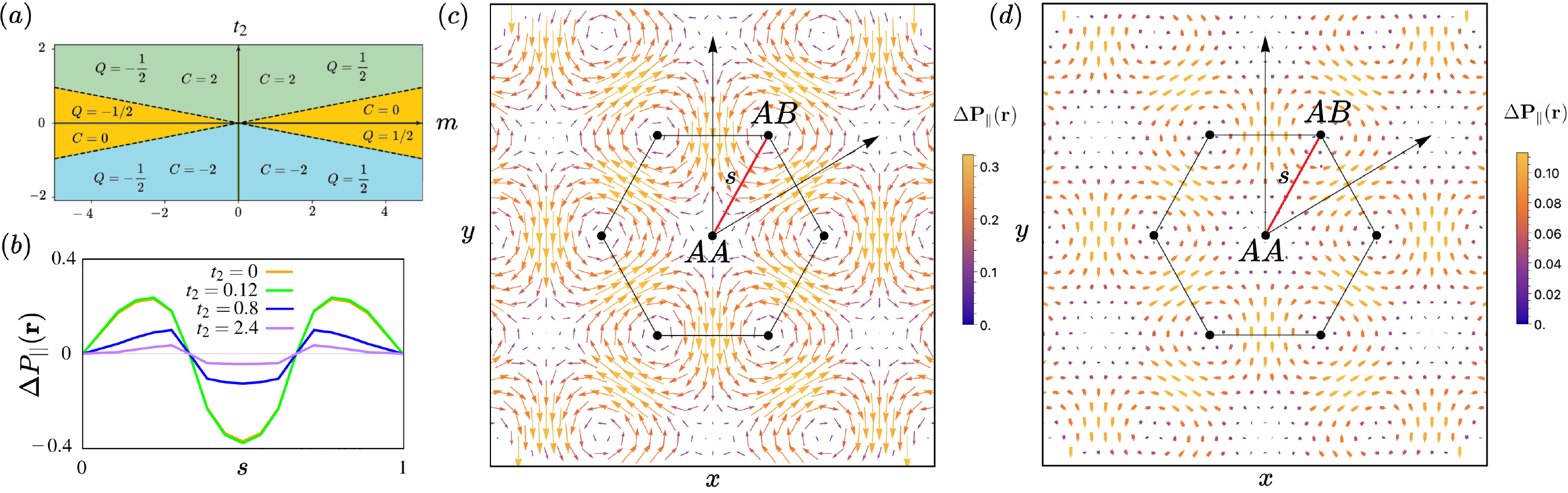}
\caption{
{\bf (a)} Phase diagram of the twisted Haldanium bilayer with topological winding numbers $Q$ of polarization textures. For $|m|<3\sqrt{3}t_2$, each of the layers acquires a Chern number, giving a total of $|C|=2$. {\bf (b)} In-plane local polarization ($\Delta P_{||}(\rv)$) along the $s$ direction connecting $AA$ and $AB$ stackings in twisted Haldanium bilayer with $\th \approx 5^\circ$, as indicated in plots {\bf (c)} and {\bf (d)}. The TPT occurs at $|t_2| \approx 0.43$, abruptly changing the magnitude of the polarization, while preserving the real-space topology of the texture. For an additional plot of $\Delta P_{||}(\rv)$ vs. $t_2$, see Appendix~\ref{app::D}. {\bf (c)}~Polarization texture in twisted Haldanium bilayer for topologically-trivial phase ($|C|=0$), realized with the parameters ${(m, t_2) = (2.25, 0)}$. The emergent polarization merons are consistent with the findings in twisted hBN bilayers~\cite{bennett2024twisted} with trivial bands.
{\bf (d)}~Twisted Haldanium bilayer with topological bands ($|C|=2$), at the point ${(m, t_2) = (2.25, 0.8)}$ of the phase diagram. The polarization texture preserves the winding, but the magnitude of the local polarization is significantly reduced across the TPT, upon the addition of next-nearest neighbor hoppings which experience staggered magnetic fluxes. The local polarization was expressed in the units of the quantum of polarization~\cite{bennett2023theory}. 
}
\label{Fig3}
\end{figure*}
As shown in Fig.~\ref{Fig3}, by tuning the Haldane mass~\cite{haldane1988model} ($t_2$) and the Semenoff mass~\cite{PhysRevLett.53.2449} ($m$), we find that the nontrivial band topology can modify the polarization texture. Correspondingly, the local polarization constituting the polarization texture can be discontinuously reduced, while preserving the topological character, i.e., the winding number. In other words; despite a significant change in the magnitude of the polarization across the TPT, the vorticity of the polarization texture is preserved. 
Here, the winding $Q$ of the polarization texture is given by~\cite{bennett2023polar}
\beq{eq:winding}
Q = \frac{1}{4\pi}\int \hat{\P} (\vec{x}) \cdot \Big( \partial_x\hat{\P} (\vec{x}) \times \partial_y\hat{\P}(\vec{x}) \Big) ~\dd^2\vec{x}
\eec
where $\hat{\P}$ is the normalized local polarization, and the integration is performed over an individual polar domain. Correspondingly, $Q=\pm 1/2$ indicates a presence of the merons/antimerons in the triangular domain spanned between $AA$ stacking points.
Instead, it should be noted that rather than trivializing the merons across TPTs, such real-space topological polarization features survive discontinuous jumps, and are retrieved across a~metallic critical point. In particular, at every stacking configuration, apart from the non-polar \textit{AA}, where the local polarization is always identically zero, the polarization approximately retains its direction respecting the stacking geometry, see~Fig.~\ref{Fig3}. 
\section{Discussion}
Our findings, supported by analytical arguments and numerical model validation, not only offer a well-defined way of capturing local polarization in crystal supercells with topological bands, but also provide a natural connection to the band topology of the supercell, while also going beyond the configuration space approximation used in the previous works \cite{bennett2023polar,bennett2023theory}. The localization of hybrid Wannier functions central to our method provides a very natural picture and intuition for the polarization components as sums of hybrid Wannier charge center (HWCC) contributions. While in a real material calculation, it might be more convenient to use the definition using Born charges within density functional perturbation theory (DFPT), the language of the hybrid Wannier functions directly indicates the band topology central to this work through the winding of SHPs, unlike the Born charges. Contrary to the previous formulation of the local polarization in terms of the local configuration-dependent Berry phase~\cite{bennett2023polar,bennett2023theory}, which requires a full momentum-space integration to obtain the local polarization, our definition is in a sense more local, with one momentum-space integration excluded by introducing HWCCs, which further supports the intuition behind our definition, conceptually consistent with the notion of locality. On the other hand, we stress that it is not possible to completely exclude the momentum-space integration, as under a full Wannierization, the non-trivial band topology requires that the Wannier functions are not exponentially localized, and Wannier centers are ill-defined, contrary to the HWCCs. Additionally, we stress that despite the reference to the notion of a local configuration, the computation of local polarization $P(\rv_j)$ or SHP $P^h_\b(\rv_j, \kperp)$ does not require the configuration space approximation. 
This is a crucial distinction, given that the topological states obtained under such approximations might be Wannierizable in configuration space, despite the non-Wannierizability of the minibands in real space.
In particular, such a scenario arguably occurs in twisted TMDs such as $t$-MoTe$_2$ with Chern bands, while the commensurate homobilayers MoTe$_2$ are deemed topologically trivial (in the 1T$^\prime$ phase) \cite{proofs2A14ref}. As we show, our formulations do not suffer from such kind of ambiguities, and furthermore allow to explicitly study TPTs which may occur in crystal supercells. While the links between topological phase transitions and associated changes in polarization captured by the geometric Berry phases according to the modern theory of polarization \cite{king1993theory, Sinisa2009} have been established in simple systems without supercells \cite{Yoshida_2023_0, Yoshida_2023}, we report an analogous effect in crystal supercells, e.g.~provided by polar heterostructures supporting topologically nontrivial polarization textures in real space. 
It is important to note that for the other types of band topologies, i.e.~upon the inclusion of additional symmetries, such as time-reversal in quantum spin Hall insulators, the bands are completely Wannierizable, if a gauge is chosen to $not$ respect the symmetry protecting the invariant. 
However, if a gauge satisfying the symmetry is chosen, the non-Wannierizability issue for defining the local polarization can be tackled similarly to the framework proposed here for the Chernful supercells. 
The study of polarization textures in the context of other band topologies is left as a subject of future research.

In the context of the topology of Chernful supercells, it~should be stressed that our definition of $\Delta C (\{\rv_j\})$, Eq.~\eqref{eq:C-loc}, captures how the {\it total} Chern number changes with respect to {\it local} perturbations in the individual parts of a supercell.
It is naturally quantized, quantifying changes of the total anomalous Hall response of an insulating supercell, and hence is well-defined. 
We note that such quantum electronic transitions, as induced in the presence of a superlattice potential, may be of technological interest, given that it shows that the Chern topology, partial or local in the form of a domain in a supercell, can be controlled with an external potential, thus changing the anomalous Hall conductivity \textit{locally}.
A change in the magnitude of the polarization texture is associated with this type of trivialization.
This is consistent with the finding of the polarization jumps on trivializing topology by changing the Hamiltonian parameters in the Haldane model without a superlattice potential~\cite{Yoshida_2023}.

Finally, we note that our findings are not limited to the Haldane model, but are expected in any Chern insulator with additional supercell length scales and with local inversion symmetry breaking. It should be noted, however, that the Haldane model is of particular relevance for the real materials, and was realized experimentally in monolayer hBN~\cite{Mitra2024} most recently. Therefore, the polar twisted Haldanium heterostructures considered in this work can be in principle engineered in real material setup. Furthermore, we note that the presence of additional symmetries such as time reversal~\cite{KaneMele} may lead to invariants beyond Chern numbers, as captured by the tenfold way ~\cite{Kitaev, SchnyderClass}, or by further taking into account the role of crystalline symmetries~\cite{Clas1, Slager2013, Slager2017, Po2017, Slager2019, Bradlyn2017}, possibly culminating in multi-gap topologies  ~\cite{bouhon2020geometric, BJY_nielsen, bouhon2019nonabelian, slager2022floquet, Jiang2021, Ahn2019,proofs1A10ref,proofs1A12ref}. The interplay of such symmetries within the above context of polarization textures presents indeed an interesting future pursuit in itself.

\section{Conclusions}
In this work, we show how local polarization textures can be defined in crystal supercells with topologically nontrivial bands.
We introduce the concept of semilocal hybrid polarization (SHP), the winding of which captures the quantized Chern numbers across TPTs within supercells. We demonstrate our findings with models for Chern insulators under superlattice potentials imposed externally, or internally, by an adequate stacking of a~moir\'e structure.
We verify these concepts using two examples, namely a Chern insulator in a superlattice potential, and two Chern insulators with a stacking mismatch, forming a moir\'e superlattice.
By calculating the polarization textures on both sides of a TPT, we find that the magnitude of the local polarization decreases when going from a trivial to a nontrivial phase, but it does not vanish completely.
Our findings show that local polarization textures may persist in systems with nontrivial band topology, and that band topology and polar topology in real space may coexist.

Additionally, we show that one can change the band topology of a supercell, purely by the local perturbations imposed on its subsystems. As a consequence, one could also control the presence of associated edge currents by the use of an external superlattice potential combined with local probes, which may be of interest for applications of novel electronics involving Chern insulators. 

Our theoretical results are of relevance for real polar materials with Chern bands, such as twisted MoTe$_2$ heterostructures. 
Engineering the parameter tuning to manipulate polarization textures with external superlattice potentials, or within moir\'e materials with nontrivial band topology, may be of potential use in optical or electronic devices. Finally, our theoretical framework is generalizable to other topological multilayers. 
\begin{acknowledgements}
    W.~J.~J. and R.-J.~S. thank Shuichi Murakami for helpful discussions. 
    W.~J.~J. acknowledges funding from the Rod Smallwood Studentship at Trinity College, Cambridge. 
    D.~B.~acknowledges the US Army Research Office (ARO) MURI project under grant No.~W911NF-21-0147 and the Simons Foundation award No.~896626. 
    R.-J.~S. and G.~C.~acknowledge funding from a New Investigator Award, EPSRC grant EP/W00187X/1. 
    R.-J.S. also acknowledges funding from a EPSRC ERC underwrite grant EP/X025829/1 as well as Trinity College, Cambridge.
\end{acknowledgements}

\appendix

\section{Local polarization in Wannierizable supercells}\label{app::A}
Here, we review the formalism of gauge-invariant local polarization in crystal superlattices, introduced in our previous work \cite{bennett2023theory}.
We start with the equivalent definitions in terms of Wannier charge centers (WCC) and Born effective charges. Both definitions were crucial for studying the local polarization in moir\'e polar heterostructures \cite{bennett2022electrically,bennett2022theory,bennett2023theory,bennett2023polar}, and were based on the introduction of the local displacements $\x(\rv_j) = \{x_{\k,\a}\}$, which correspond to perturbing atoms $\k$ in the directions $\a$. Accordingly, for the local polarization $\P(\rv_j)$ in the unit cell at $\rv_j$, we could write
\beq{eq:P-wannier-modern}
\P(\rv_j) = -\frac{ef}{\Om_0}\sum_{n}^{\occ} \int_0^{\x(\rv_j)} \partial_{x'_{\k,\a}}\bar{\w}_n \dd x'_{\k,\a}
\eec
where the Einstein summation convention for the indices $\k$ and $\a$ was used, $n \in \text{occ}$ are the band indices of occupied bands, and WCC are defined as ${\bar{\w}_n \equiv \braket{w_{n,0}|\hat{\rv}|w_{n,0}}}$ in terms of the Wannier functions represented by the states \cite{vanderbilt1993electric,king1993theory,marzari1997,souza2001,marzari2012maximally},
\beq{Wannier}
    \ket{w_{n,\R}} = \frac{\Om_{\text{sc}}}{(2\pi)^3}\oint_{\scBZ}e^{-i\kv\cdot\R}\ket{\psi_{n,\kv}}\dd\kv.
\eeq
Here, $\Om_{\text{sc}}$ denotes the real-space supercell volume, $\scBZ$ is the corresponding Brillouin zone associated with the superlattice, and $\R$ is a supercell position vector. Equivalently, we can express the local polarization in an alternative form, using dynamical Born effective charges. Componentwise, it reads
\beq{eq:P-approx-2}
P_{\b}(\rv_j) = \frac{1}{\Om_{0}}\int_0^{\x(\rv_j)}Z^*_{\k,\a\b}(\x')\dd x'_{\k,\a}
\eec
with Born charges defined as $Z^{*}_{\k,\a\b} = \Omsc \pd{P_{\b}}{x_{\k,\a}}$, which in terms of the bands and phonon displacements of atoms $\kappa$ in direction $a$, $x_{\k,\a}$, can be expressed as \cite{gonze1997dynamical,ghosez1998dynamical}
\beq{eq:Z-mixed}
Z^{*}_{\k,\a\b}(\x(\rv_j)) = \frac{-2ief\Omsc}{(2\pi)^3}\sum_{n}^{\occ} \oint_{\scBZ} \braket{\partial_{x_{\k,\a}}u_{n,\kv}|\partial_{k_{\b}}u_{n,\kv}}\dd\kv.
\eeq
Hence, consistently with the previous expression, in the Wannierizable systems, we retrieve,
\beq{Born_equiv}
    Z^{*}_{\k,\a\b}(\x) = -\frac{ef}{\Om_0}\sum_{n}^{\occ} \partial_{x_{\k,\a}} \bar{\w}_{n,\beta}.
\eeq{}
Here, $\Om_{0}$ is the volume of a unit cell, with $\x(\rv_j)$ corresponding to a set of displacements of cores in unit cell $\rv_j$.
Last, we note that in the context of real materials, the Born charge definition can be naturally extended by the use of the nonadiabatic Born effective charges (NABECs) introduced in Ref.~\cite{Dreyer2022}. Here, under the implementation of NABECs to deduce the local polarization at $\rv_j$, an analogous integration to the one adapted for the regular Born charges in Ref.~\cite{bennett2023theory} could be performed, i.e. 
\beq{eq:P-Born}
    P_\beta(\vec{r}_j)= \frac{1}{\Om} \int^{\x(\vec{r}_j)}_0 Z^{*}_{\k,\a\b}(\om \rightarrow 0,~\x(\vec{r}_j)) \dd x_{\k,\a},
\eeq
which for insulators, in the $\om \rightarrow 0$ limit, coincides with the Eq.~\eqref{eq:P-approx-2}. Here, the NABECs at the frequency $\om$ are given by $Z_{\k,\a\b}^{*}(\om)$~\cite{Dreyer2022} in the context of a two-dimensional, as relevant to this work, topologically nontrivial system, amounting to
\begin{align}\label{eq:P-Born}
    & Z^{*}_{\k,\a\b}(\om) = -\text{Im}~\text{lim}_{\eta \rightarrow 0^+} \int_{\text{BZ}} \frac{\dd^2 \kv}{(2\pi)^2}  \sum_{n \neq m} \frac{f_{n\kv}-f_{m\kv}}{E_{n\kv}-E_{m\kv}+\om+i\eta}\notag\\ 
    & \hspace{2cm} \times \braket{u_{n,\kv}|  \partial_{k_a} u_{m,\kv}}\bra{u_{m,\kv}} \partial_{x_{\k,\a}} H \ket{u_{n,\kv}}.
\end{align}
To obtain the NABECs, the bands $\ket{u_{n,\kv}}$ with energies $E_{n\kv}$ and Fermi-Dirac occupation factors $f_{n\kv}$ are used, and the derivatives of the Hamiltonian subject to the local phonon displacements $\partial_{x_{\k,\a}} H$ are evaluated. It should be noted that here, as in the rest of the work, only the electronic contribution to the polarization is considered, while the ion (core) contribution (which trivially obtains a dipole moment of core charges) is not included. 

\section{Semilocal hybrid polarization}\label{app::B}
\begin{figure}
      \includegraphics[width=\columnwidth]{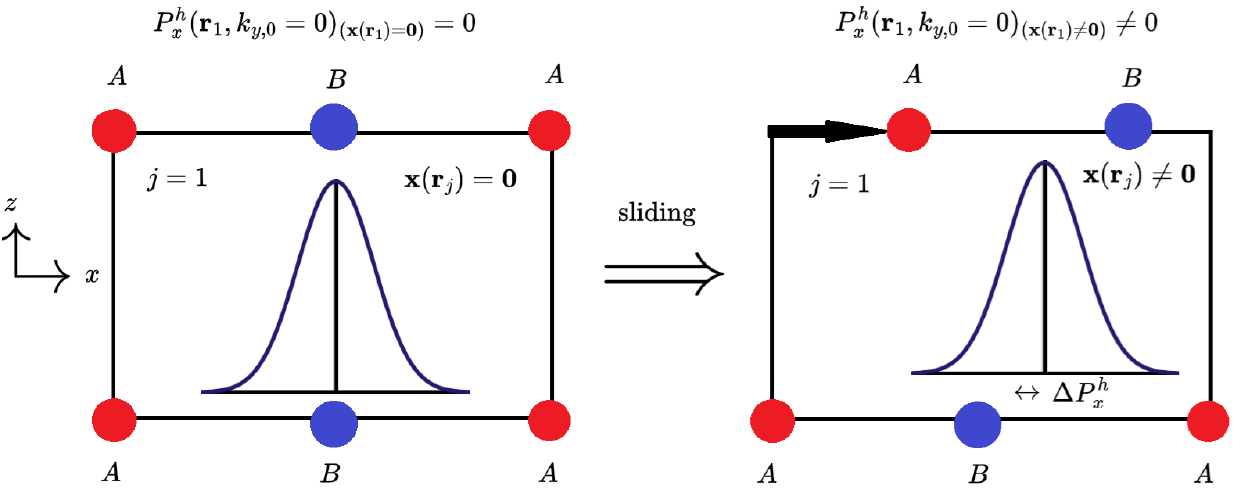}
      \caption{\textbf{(a)} HWCC in the reference configuration ($\vec{x}(\rv_j) = \textbf{0}$) under the consistently choosen gauge (fixing $\vec{k}_0$).  \textbf{(b)} Flow of the HWCC subject to the sliding ($\vec{x}(\rv_j) \neq \textbf{0}$), which allows to reconstruct the local polarization in a subcell of a given configuration.}
\label{FigS1}      
\end{figure}

In this section, we extend our definition of local polarization to non-Wannierizable systems, such as Chern insulators, in further detail. As Wannier functions are not exponentially localized in such cases, the polarization in terms of WCC is ill-defined for arbitrary gauges \cite{Sinisa2009, Yoshida_2023}. However, hybrid Wannier functions (HWFs) exponentially-localized in \textit{one} direction, which we denote as $||$, can be defined,
\beq{hWannier}
    \ket{w^h_{n,\vec{R}}(\kperp)} = \frac{\Om_{\text{sc}}}{(2\pi)^2}\oint_{\scBZ} e^{-i k_{||}  (\vec{R})_{||}}\ket{\psi_{n,\kv}}~\dd k_{||}.
\eeq
Here, $(\vec{R})_{||}$ is the component of a superlattice vector $\R$, which is parallel to the direction, in which the polarization component is to be deduced. Analogously, the hybrid Wannier charge centers (HWCC) can be introduced. On introducing a shortcut notation, $\ket{w^h_{n}(\kperp)} \equiv \ket{w^h_{n,\vec{0}}(\kperp)}$, we have,
\beq{HWCC}
    \bar{w}^h_{n,\beta}(\kperp) \equiv \braket{w^h_{n}(\kperp)|\hat{r}_{\beta}|w^h_{n}(\kperp)},\\
\eeq
with $\hat{r}_{\beta}$, the position operator components. Before defining the semilocal version of the hybrid polarization with the introduced HWCCs, we define the $total$ hybrid polarization itself. Componentwise, it reads,
\beq{}
    P^h_\beta(\kperp) = -\frac{ef}{\Om_{\text{sc}}} \sum^{\text{occ}}_n \bar{w}^h_{n,\beta}(\kperp).
\eeq
On introducing the notion of local configuration for defining the local polarization, consistently with Ref.~\cite{bennett2023theory}, we can now define the semilocal hybrid polarizations (SHPs), see also Fig.~\ref{FigS1}. Namely, using phonon displacements, or equivalently, depolarizing perturbations (e.g. in the super-Haldane model---equivalent to setting the vanishing onsite potential), which directly encode the local configuration $\x(\rv_j)$, we write,
\beq{eq:P-hwannier-loc}
P^h_\beta(\rv_j, \kperp) = -\frac{ef}{\Om_0}\sum_{n}^{\occ} \int_0^{\x(\rv_j)} \partial_{x'_{\k,\a}}\bar{w}^h_n(\kperp) ~\dd x'_{\k,\a}.
\eeq
It should be noted that, with $P^h_\beta(\rv_j, \kperp)$ introduced as a change in the gauge-invariant sum of the HWCCs over occupied band indices (or equivalently, change of the Berry phase), the SHPs are definitionally gauge-invariant objects.
Furthermore, we know that physically, on adding up electric dipole moments associated with local polarizations, one obtains the total polarization $\textbf{P}_{\text{tot}}$,
\beq{}
    \textbf{P}_{\text{tot}} = \frac{1}{N_\text{tot}} \sum^{N_\text{tot}}_j \vec{P}(\rv_j),
\eeq
which is also consistent with the additivity of phonon displacements in the integral limits $\x(\vec{r}_j)$. $N_\text{tot}$ is the total number of subcells contained in a supercell ($N_\text{tot} = N_x N_y$, for two spatial dimensions). By an analogous argument, we have
\beq{}
    P^h_{\beta}(k_{\perp}) = \frac{1}{N_\text{tot}} \sum^{N_\text{tot}}_j P^h_{\beta}(\rv_j, k_{\perp}).
\eeq
which, on summing over $k_\perp$ as detailed in the main text, provides an adequate decomposition of the chosen total polarization component into contributions associated with distinct unit subcells. Here, it should be noted that all polarizations in this work considered under periodic boundary conditions, are defined modulo a quantum of polarization respecting the superlattice vector $\textbf{R}$. In the context of local polarizations within crystal supercells, such modular character, intrinsically due to the gauge ambiguity, was in fact discussed in detail in Ref.~\cite{bennett2023theory}.
Next, we remark on the relations between the hybrid polarizations (or equivalently HWCCs), the polarization currents $\vec{j}_P = \vec{\dot{P}}$, and Berry phases $\phi_n(\kperp)$ defined in the main text, which further justify the proposed construction of the local polarization definition utilizing SHPs. Within an independent-particle picture, the time-dependent polarization current can generally be decomposed in a two-dimensional system into individual contributions as~\cite{vanderbilt2018berry, resta2022notes},
\beq{}
    \vec{j}_P(t) = f \sum^{\text{occ}}_{n} \int \frac{\dd^2 \textbf{k}}{(2\pi)^2}~  \vec{j}_{n\kv}(t),
\eeq
where $f$ is the occupation factor in the valence bands ({${f=2}$ for spin-degenerate systems in zero-temperature limit}). In~terms of the individual contributions under an adiabatic current-inducing perturbation, we obtain to first order
\begin{align}\label{eq:ind}
    & \vec{j}_{n\kv}(t) = -e \bra{u_{n,\kv}} \hat{\vec{v}} \ket{u_{n,\kv}}\notag\\ 
    & \hspace{1cm } - ie\hbar \sum_{m \neq n} \Big[ \frac{\braket{\dot{u}_{n,\kv}| u_{m,\kv}} \bra{u_{m,\kv}} \hat{\vec{v}} \ket{u_{n,\kv}}}{E_{n\kv} - E_{m\kv}} - \text{c.c.} \Big],
\end{align}
where $\hat{\vec{v}} = \frac{1}{\hbar} \nabla_\kv H(\kv)$ is the velocity operator, and $H(\kv)$ is the Bloch Hamiltonian. On further recognizing that for $n \neq m$, $\braket{u_{n,\kv}|\nabla_\kv u_{m,\kv}} = \frac{\bra{u_{n,\kv}} \nabla_\kv H \ket{u_{m,\kv}}}{E_{m\kv} - E_{n\kv}}$, one obtains,
\beq{}
    \vec{j}_{n\kv}(t) = -e \bra{u_{n,\kv}} \hat{\vec{v}} \ket{u_{n,\kv}} - ie \Big[ \braket{\dot{u}_{n,\kv}|\nabla_\kv u_{n,\kv}} - \text{c.c.}
    \Big].
\eeq
Furthermore, on substituting to $\Delta \textbf{P} \equiv \int \textbf{j}_\text{P} \dd \text{t}$, which was introduced in the main text, the Eq.~\eqref{eq:ind} results in
\beq{}
\Delta P_{\beta} = -ef  \int^1_0 \dd \lambda \sum^{\text{occ}}_{n} \int \frac{\dd^2 \textbf{k}}{(2\pi)^2} i\Big[ \braket{\partial_\lambda u_{n,\kv}|\partial_{k_\beta} u_{n,\kv}} - \text{c.c.}
    \Big], 
\eeq
consistently with the seminal formula of~Ref.~\cite{vanderbilt1993electric}. Here, the variables were changed from $t$ to $\lambda$, with $\la$ parametrizing the time-dependent adiabatic switching of the perturbation which induces the polarization. We quote the general result from Ref.~\cite{vanderbilt2018berry}: $\bar{w}^h_{n,\beta}(\kperp) = \frac{L_\beta}{2\pi} \phi_n(\kperp)$, which we combine with Eq.~\eqref{eq:Berry}, $\phi_n(k_\g) \equiv i \oint_{\kperp = k_\gamma} \braket{u_{n,\kv}|\partial_{k_\beta} u_{n,\kv}} \dd k_\beta$, and a choice of the adiabatic perturbations $\la \equiv \{x_{\k,\a}\}$, equivalent to the local displacements $\vec{x}(\rv_j)$. Upon direct insertion of these identities, we~finally obtain Eq.~\eqref{P-local},
\beq{}
    \Delta P_{\beta}(\rv_j) =  -\frac{ef}{\Om_0}\sum_{n}^{\occ} \oint_{\mathcal{C}} \int_0^{\x(\rv_j)} \partial_{x'_{\k,\a}}\bar{w}^h_{n,\b}(\kperp)\dd x'_{\k,\a} \dd\kperp.
\eeq
This concludes the derivation, which we include to expose the correspondences between polarization currents, Berry phases, and HWCCs that were used to define the hybrid polarizations. Manifestly, we note that the Eq.~\eqref{P-local}, which directly captures the local polarization in the supercells with Chern bands, can be partitioned into the semilocal hybrid polarizations, as was explicitly presented in the Eq.~\eqref{eq:P-local} of the main text.

\section{Conventional Chern markers}\label{app::C}
Importantly, a superlattice potential (or even more generically, a random potential disorder, which defines a supercell of infinite size) changes/removes the periodicity of the system. In the case of systems realizing crystalline superlattices, the topological invariants may become computationally costly to evaluate, or in the latter case, may be no longer possible to deduce as $\kv$-integrals over a well-defined BZ. Therefore, under the settings of such kinds: local, real-space indicators (markers) for bulk topology are in demand.

For the Chern topology central to this work, the Chern markers satisfying these conditions can be defined \cite{Kitaev20062, Bianco_2011, PhysRevB.109.014206} under both periodic and open boundary conditions. To achieve this goal, we follow the derivation by Bianco and Resta~\cite{Bianco_2011}, that starts by recognizing that
\begin{equation}
  C = -\frac{1}{\pi} \mathfrak{Im} \sum^{N_\text{occ}}_{n=1} \sum^{\infty}_{m = N_{\text{occ}+1}} \int_{\textnormal{BZ}} \dd^2 \textbf{k}~ \braket{\partial _{k_x} u_{n,\kv} | u_{m,\kv}}\braket{u_{m,\kv} | \partial _{k_y} u_{n,\kv}},  
\end{equation}
where $N_\text{occ}$ denotes the number of occupied bands, and on substituting the identity for $n \neq m$, with $\ket{\psi_{n,\textbf{k}}} = e^{i\kv \cdot \rv} \ket{u_{n,\kv}}$,
\begin{equation}
  \braket{u_{n,\kv} | \nabla_\textbf{k} u_{m,\kv}} = -i \braket{\psi_{n,\textbf{k}} | \hat{\textbf{r}}|\psi_{m,\textbf{k}}}, 
\end{equation}
the expression for the Chern number can be rearranged into
\begin{align}
  & C = -\frac{A}{4\pi^3} \mathfrak{Im} \sum^{N_{\text{occ}}}_{n=1} \sum^{\infty}_{m = N_{\text{occ}}+1} \int_{\textnormal{BZ}} \dd^2 \textbf{k} \dd^2 \textbf{k}^{\prime} \notag\\ 
  & \hspace{2cm}\times \braket{ \psi_{n,\textbf{k}} |\hat{x}| \psi_{m,\textbf{k}^{\prime}}}\braket{\psi_{m,\textbf{k}^{\prime}} |\hat{y}| \psi_{n,\textbf{k}}}.
\end{align}
Here, $A$ represents the unit cell area of the system, and the vanishing of the matrix elements for $\textbf{k}^{\prime} \neq \textbf{k}$ is exploited. On defining the projectors onto occupied and unoccupied states ($\hat{P} + \hat{Q} = 1$), which can be written as
\begin{equation}
    \hat{P} = \frac{A}{(2\pi)^2} \sum^{N_\text{occ}}_{n=1}  \int_{\textnormal{BZ}} \dd^2 \textbf{k} ~ \ket{\psi_{n,\textbf{k}} }\bra{\psi_{n,\textbf{k}}}, 
\end{equation}
\begin{equation}
    \hat{Q} = \frac{A}{(2\pi)^2} \sum^{\infty}_{m = N_{\text{occ}}+1} \int_{\textnormal{BZ}} \dd^2 \textbf{k}^\prime ~ \ket{\psi_{m,\textbf{k}^\prime} }\bra{\psi_{m,\textbf{k}^\prime}} ,
\end{equation}
one finally arrives at the Chern marker formula, after having inserted the resolution of the identity in the localized orbital basis ($1 = \sum_{j,a} \ket{\vec{r}_{j,a}} \bra{\vec{r}_{j,a}}$),
\beq{eq::Cmarker}
    C(\vec{r}_j) = -\frac{4\pi}{A} \mathfrak{Im} \sum_{\textnormal{a} \in \textnormal{cell}} \braket{\textbf{r}_{j,a} |\hat{P}\hat{x}\hat{Q}\hat{y}| \textbf{r}_{j,a}} \equiv \frac{4\pi}{A} \mathfrak{Im}~ \textnormal{Tr}_{\textnormal{cell}} \{\hat{P}\hat{x}\hat{P}\hat{y}\}.
\eeq
In particular, under periodic boundary conditions, ${C = \frac{1}{N_{\text{tot}}} \sum_j C(\rv_j)}$, which in a continuum limit can be written as ${C = \frac{1}{A_\text{tot}} \int \dd^2 \rv~ C(\rv_j)}$, with $A_\text{tot} = \int \dd^2 \rv$. Numerically, one can average Chern markers over multiple unit cells, obtaining local Chern numbers (LCN), which should converge to the Chern numbers on inclusion of a sufficient number of cells. We adapt an implementation of the projectors used for evaluating the markers, consistently with Ref. \cite{PhysRevResearch.2.013229}. It should be noted that while definition Eq.~\eqref{eq::Cmarker} appears to be ill-behaved in systems under periodic boundary conditions, which we consider in this work, it can be manifestly recast in a well-defined way~\cite{PhysRevB.109.014206},
\beq{}
        C(\vec{r}_j) = -\frac{4\pi}{A} \mathfrak{Im} \sum_{\textnormal{a} \in \textnormal{cell}} \braket{\textbf{r}_{j,a} |\hat{P}[\hat{x},\hat{P}][\hat{y},\hat{P}]| \textbf{r}_{j,a}},
\eeq
where it is recognized that the commutators $[\hat{x},\hat{P}]$ and $[\hat{y},\hat{P}]$ are well-behaved. Ultimately, for the evaluation included in Fig.~\ref{Fig2}(e), we apply the Chern markers under open boundary conditions, with the superpotential applied to a bulk subsystem of size $51 \times 51$ subcells, within a slab of size $71 \times 71$. The boundary of the $71 \times 71$ system hosts values of the marker, which cancel the bulk contributions after a complete summation over the entire system, consistently with the general expectation of the method~\cite{Bianco_2011}.

\section{Effective models for twisted Haldanium bilayer}\label{app::D}

We here elaborate on the models for twisted Haldanium bilayer realizing topological bands. First, we introduce an effective tight-binding (TB) model, which is based on the configuration space approximation picture~\cite{bennett2023polar}. Finally, we conclude by introducing a continuum Bistritzer-MacDonald (BM) model for the low-energy physics of the model.

\subsection{Tight binding model}

The local polarization in moir\'e bilayers can be obtained from a simple TB model, as shown in~Ref.~\cite{Yu_2023}. However, the simplicity comes at the cost of the configuration space approximation. In the case of moir\'e hBN bilayers, effectively, one can model such insulators by considering four bands $(\ket{u_{v_t,\kv}}, \ket{u_{v_b,\kv}}, \ket{u_{c_t,\kv}}, \ket{u_{c_b,\kv}})$, i.e.~valence and conduction bands of two uncoupled layers. Following the Ref.~\cite{Yu_2023}, the monolayer gap can be approximated as $\Delta E_\kv = E_{c_t,\kv} - E_{v_t,\kv} \approx E_{c_b,\kv} - E_{v_b,\kv}$, and the interlayer tunneling can be treated perturbatively, to second order, hybridizing bands as
\beq{}
    \ket{\tilde{u}_{v_{t/b},\kv}} \approx \Big( 1 - \frac{1}{2} \Big| \frac{t_{v_{t/b}c_{b/t},\kv}}{\Delta E_\kv} \Big|^2 \Big) \ket{u_{v_{t/b},\kv}} - \frac{t_{v_{t/b}c_{b/t},\kv}}{\Delta E_\kv} \ket{u_{c_{b/t},\kv}},
\eeq
\beq{}
    \ket{\tilde{u}_{c_{t/b},\kv}} \approx \Big( 1 - \frac{1}{2} \Big| \frac{t_{v_{b/t}c_{t/b},\kv}}{\Delta E_\kv} \Big|^2 \Big) \ket{u_{c_{t/b},\kv}} + \frac{t^*_{v_{b/t}c_{t/b},\kv}}{\Delta E_\kv} \ket{u_{c_{b/t},\kv}}.
\eeq
Here, $t_{v_{t/b}c_{b/t},\kv}$ are matrix elements providing intergap interlayer coupling and the convention for the interlayer coupling (that satisfies the above perturbation relations) is $t_{v_{t/b}c_{b/t},\kv} = \bra{u_{c_{b/t},\kv}} H \ket{u_{v_{t/b},\kv}}$.  Due to the Fermi occupation of states, the effects of $t_{{c/v}_{t/b} {c/v}_{t/b},\kv}$ are negligible at second order. Crucially, $t_{v_{t/b}c_{b/t},\kv}$ varies between configurations described by different sliding vectors $\vec{x}$ and can be computed from a TB Hamiltonian $H_{\text{moir\'e}, \text{TB}}$:
\beq{}
    H_{\text{moir\'e}, \text{TB}} =
\begin{pmatrix}
    \tilde{m}_\kv & t_\kv & t_{A_t A_b, \kv} & t_{A_t B_b, \kv} \\
    t^*_\kv & -\tilde{m}_\kv & t_{B_t A_b, \kv}  & t_{B_t B_b, \kv}\\
    t^*_{A_t A_b, \kv} & t^*_{B_t A_b, \kv} & \tilde{m}_\kv & t_\kv\\
    t^*_{A_t B_b, \kv} &  t^*_{B_t B_b, \kv} & t^*_\kv & -\tilde{m}_\kv\\
\end{pmatrix},
\eeq
where the off-diagonal $2~\times~2$ blocks define the orbital and stacking-dependent tunneling $T(\kv, \vec{x})$. In particular, $T(\kv, \vec{x})$ corresponds to the top-right $2~\times~2$ block, while the bottom-left block constitutes $T^\dag(\kv, \vec{x})$.  Here, we label atoms in top and bottom layers as $(A_t, B_t, A_b, B_b)$ (which defines the basis of the Bloch states of the two atomic species), and contrary to Ref.~\cite{Yu_2023}, we consider a combined Semenoff $m$ and Haldane mass $t_2$ in the form of $\tilde{m}_\kv = m - 2 \sum_i t_2 \sin(\textbf{k} \cdot \vec{b}_i)$, with $\vec{b}_i$ the vectors corresponding to the second-neighbor hoppings. On the other hand, the in-plane $nearest$-neighbor hoppings read $t_\kv = \bra{u_A} H \ket{u_B} = t_1 \sum^3_{i=1} e^{i \kv \cdot \Delta \R_i}$, where $\Delta \R_i$ label nearest-neighbor displacements (from the \textit{A} to \textit{B} species according to the convention used for $t_\kv$). Out-of-plane displacements are given by $\Delta \R_{A_b A_t} = \Delta \R_{B_b B_t} = \vec{x}$, $\Delta \R_{A_b B_t, i} = \vec{x} + \Delta \R_i$, $\Delta \R_{B_b A_t, i} = \vec{x} - \Delta \R_i$. Note that here, $\vec{x}$ can be chosen to lie in the Wigner-Seitz unit cell of a \textit{B} atom. On the same order of approximation as the intra-layer nearest-neighbor coupling series truncation, it is sufficient to consider coupling with atoms connected by the aforementioned out-of-plane displacements, since only these atoms can lie inside the Wigner-Seitz unit cell for any $\vec{x}$. From solving the monolayer problem first, one obtains unperturbed bands $\ket{u_{{c/v}_{t/b},\kv}}$ in terms of $\ket{u_{A_{t/b}}}$ and $\ket{u_{B_{t/b}}}$ (the periodic parts of the \textit{A} and \textit{B} atomic monolayer Bloch orbitals),
\begin{align}
    \ket{u_{c_{t/b}, \kv}} &= c_{A\kv} \ket{u_{A_{t/b}}} + c_{B\kv} \ket{u_{B_{t/b}}}, \\
    \ket{u_{v_{t/b}, \kv}} &= (c^{*}_{B\kv}) \ket{u_{A_{t/b}}} - c_{A\kv} \ket{u_{B_{t/b}}},
\end{align}
where from a single-layer problem unperturbed by tunneling, one obtains the corresponding coefficients:
\beq{}
    c_{A\kv} = \frac{\tilde{m}_\kv+\sqrt{\tilde{m}_\kv^2+|t_\kv|^2}}{\sqrt{\Big(\tilde{m}_\kv+\sqrt{\tilde{m}_\kv^2+|t_\kv|^2}\Big)^2 + |t_\kv|^2}},
\eeq
\beq{}
    c_{B\kv} = \frac{t^*_\kv}{\sqrt{\Big(\tilde{m}_\kv+\sqrt{\tilde{m}_\kv^2+|t_\kv|^2}\Big)^2 + |t_\kv|^2}}. 
\eeq
Furthermore, these yield $t_{v_{t/b}c_{b/t},\kv}(\vec{x}) = \bra{w_{c_{b/t}}} H \ket{w_{v_{t/b}}}$, where $\ket{w_{c_{b/t}}}$ is the Wannier function obtained from $\ket{u_{c_{b/t}}}$,
\begin{align}
    & t_{v_{t}c_{b},\kv}(\vec{x}) = e^{i\kv \cdot \vec{x}} \Big[ c_{A \kv} c^{*}_{B \kv} (t_{A_b A_t}(\vec{x})-t_{B_b B_t}(\vec{x})) \notag\\
    &  + (c^{*}_{B\kv})^2 \sum^3_{i=1} \left( t_{B_b A_t,i}(\vec{x}) e^{-i\kv \cdot \Delta \R_{i}}\right) - c^2_{A\kv} \sum^3_{i=1} \left( t_{A_b B_t,i}(\vec{x}) e^{i\kv \cdot \Delta \R_{i}}\right) \Big],
\end{align}

\begin{align}
    & t_{v_{b}c_{t},\kv}(\vec{x})  = e^{-i\kv \cdot \vec{x}} \Big[ c_{A\kv} c^{*}_{B \kv} (t^{*}_{A_b A_t}(\vec{x})-t^{*}_{B_b B_t}(\vec{x}) ) \notag\\
    & + (c^{*}_{B\kv})^2 \sum^3_{i=1} \left( t^{*}_{A_b B_t,i}(\vec{x}) e^{-i\kv \cdot \Delta \R_{i}}\right) - c^2_{A\kv} \sum^3_{i=1} \left( t^{*}_{B_b A_t,i}(\vec{x}) e^{i\kv \cdot \Delta \R_{i}}\right) \Big].
\end{align}
Here, $t_{A_b B_t, i}(\vec{x}) = \bra{w_{A_b}} H \ket{w_{B_t}}$ and $A_b$ and $B_t$ are separated by $\Delta R_{A_b B_t, i}$ (as earlier, $\ket{w_{{A/B}_{b/t}}}$ is the Wannier function for $\ket{u_{{A/B}_{b/t}}}$). Writing in the explicit dependence of $t_{A_b B_t, i}(\vec{x})$ on $\vec{x}$,
\beq{}
    t_{{A}_b {B}_t, i}(\vec{x})=
        \begin{cases}
            t^0_{A_b B_t, i} e^{-|\Delta \R_{A_b B_t, i}(\vec{x})|/\lambda_{AB}} & \text{if } |\Delta \R_{A_b B_t, i}(\vec{x})| <  |\Delta \R_i|, \\
            0 & \text{otherwise},
        \end{cases}    
\eeq
with $\lambda_{AB}$ a layer-separation dependent regularization of the hopping amplitudes, reflecting the overlaps of the orbitals between which the hopping occurs. The stacking-dependent interlayer hoppings for the other three orbital flavour combinations were regularized analogously. To obtain the results in Fig.~\ref{Fig3}, the values chosen for the various tight-binding parameters are as follows: $m=2.25$, $t_1 = 2.4$, $t^0_{A_b B_t, i} = t^0_{B_b A_t, i} = 1.28$, $t^0_{A_b A_t, i} = 0.8$, $t^0_{B_b B_t, i} = 0.6$, $\lambda_{AB} = \lambda_{BA} = 1.32$, $\lambda_{AA} = 1.36$, and $\lambda_{BB} = 1.27$, providing a faithful representation of the twisted hBN bilayer~\cite{Yu_2023}, subject to an addition of the second neighbor hoppings, which experience staggered magnetic fluxes.

With the evaluated interlayer coupling constants, the perturbed bands $\ket{\tilde{u}_{v/c_{t/b},\kv}}$ can be found within the perturbation theory, as described earlier. Finally, to deduce the in-plane polarization, the Berry connection can be furthermore obtained.~Namely, the resulting Berry connection in the calculated bands reads
\\
\beq{}
\begin{split}
    A_{cc/vv}(\kv) = - i \braket{\tilde{u}_{{c/v}_{t/b},\kv}| \nabla_\kv \tilde{u}_{{c/v}_{t/b},\kv}},
\end{split}
\eeq
which can be further integrated over $\kv$-space to obtain the Berry phase and polarization, consistently with the main text, Eqs.~\eqref{P-local},~\eqref{eq:Berry}.
Crucially, our model goes beyond the configuration space tight-binding adaptation of Ref.~\cite{Yu_2023}. In our effective model, the hoppings, which are stacking-dependent, are regularized by $both$ out-of-plane and in-plane distances. This modification more accurately reflects an overlap of the corresponding basis orbitals, which electrons experience on hopping.
\begin{figure}
      \includegraphics[width=\columnwidth]{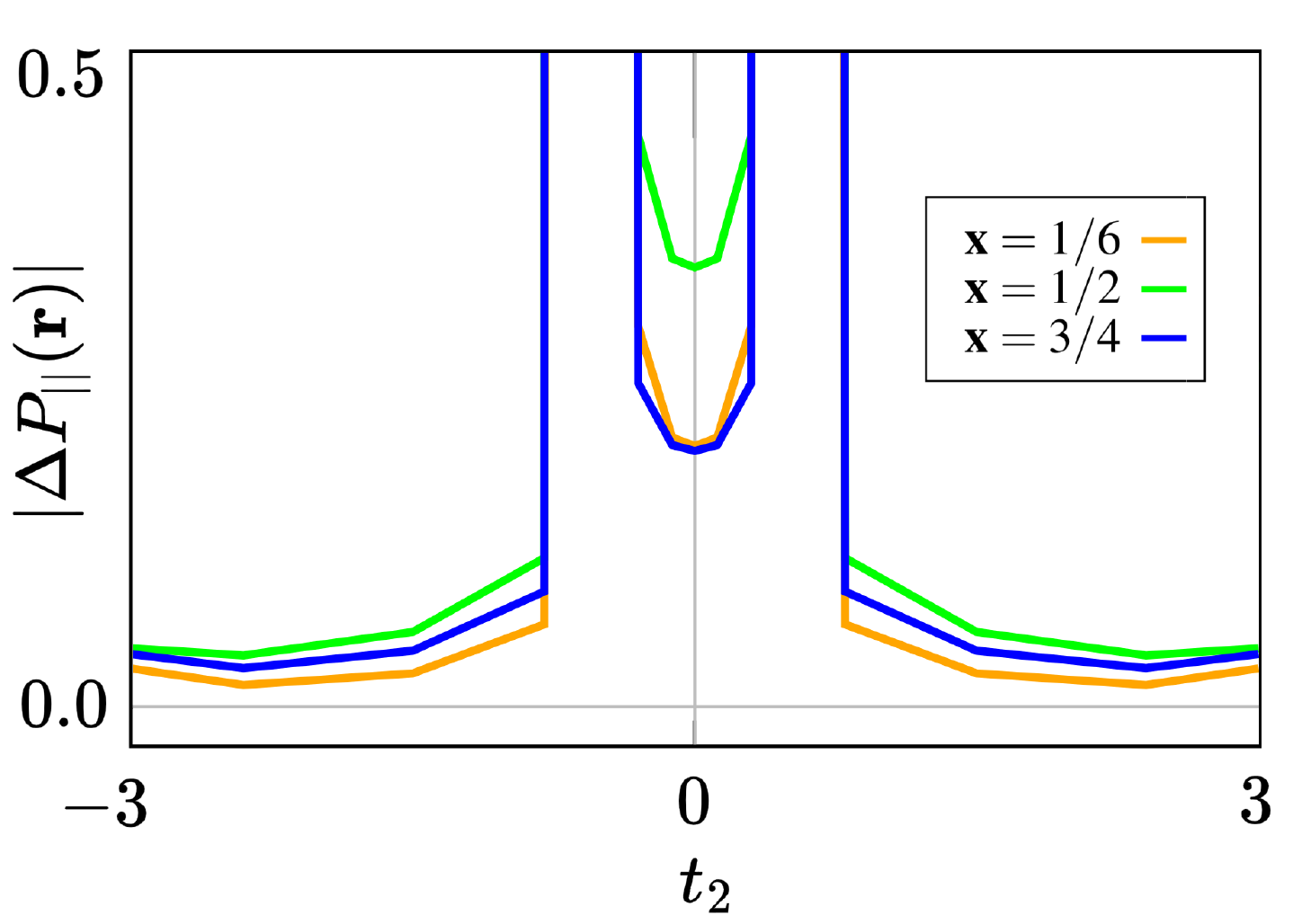}
      \caption{The in-plane local polarization $\Delta P_{||}(\rv)$ at different stackings $\vec{x}$ as a function of $t_2$. The chosen stackings $\vec{x} \equiv \vec{x}(\vec{r}) = \{1/6, 1/2, 3/4\}$ correspond to the fractional coordinates along the $s$ direction, with $s=0$ and $s=1$ corresponding to the AA bilayer stacking, see Fig.~\ref{Fig3} for further reference. We emphasize that the model breaks down in the proximity of the critical points $(|t_2| \approx 0.4)$, where the gap becomes too small for the second order perturbation theory to yield reliable values of the local electric polarization.}
\label{FigS2}      
\end{figure}
For completeness, we accordingly include a plot (Fig.~\ref{FigS2}) of the in-plane local polarization $\Delta P_{||}(\rv)$ as a function of $t_2$, which was realized and computed within the model described above. We show $\Delta P_{||}(\rv)$ for different stackings $\x$, that were defined along the direction $s$; see also Fig.~\ref{Fig3} for reference. We note that close to $t_2 = 0.43$ corresponding to the metallic critical point, the local polarization is ill-defined and diverges, as the perturbative model considered here breaks down for small gaps $\Delta E_\kv$. On the contrary, Figs.~\ref{Fig3}(c) and \ref{Fig3}(d) were obtained on both sides away from the critical point, where the gap is well-preserved.

\subsection{Continuum model}
Last, we elaborate on the continuum model for topological fermions, which captures the low-energy physics, including the electric polarization, in the twisted Haldanium bilayer. Following the continuum formulation of the local polarization introduced in our previous work Ref.~\cite{bennett2023theory}, we start by recognizing that, if the two layers are twisted, with $\th$ denoting the angle of the twist, this change can be described with a deformation field given by
\beq{eq:twist_deform}
    \Df_{\text{t}}(\rv) = -\Df_{\text{b}}(\rv) = \frac{\th}{2} \mathbf{\hat{z}} \times \rv. 
\eeq
Here, each layer experiences a deformation field $\Df_l$, and ${l= \text{t,b}}$ is the layer index. As a result of a small deformation in each individual layer, the $\psi$ electron field, introduced in the main text, is correspondingly modified as~\cite{Balents2019}
\beq{Eq:psi_deform}
    \psi_l(\rv) = \lb 1-\bm{\nabla}\cdot\Df_l(\rv) \rb^{1/2} \psi_l(\x(\rv))e^{-i\vec{K}\cdot \Df_l(\rv)}
\eec
in addition to the consistent transformation of the integration measure. Here, $\vec{K}$ is the momentum at the Dirac point, corresponding to an individual valley.
The continuum Hamiltonian of the decoupled topological bilayers can be obtained as~\cite{Balents2019}
\begin{align}\label{eq:Ham_BL}
    & H_{\text{BL}} = \sum_{l= \text{t,b}}\int \psi^{\dagger}_l \Bigg[\biggl(m + b \Big(\partial_{r_{\beta}}+\frac{\partial \Df_{l,\gamma}}{\partial r_\beta}\partial_{r_{\gamma}}\Big)^2 \biggl) \tau_3 \notag\\
    &  - iv \lb \tau^{\beta} + \frac{\partial \Df_{l,\beta}}{\partial r_\gamma}\tau^{\gamma} \rb \partial_{r_{\beta}}  + v (\vec{K}\cdot \partial_{r_\beta} \Df_{l}) \tau^{\beta}\Bigg] \psi_l \dd^2\rv,
\end{align}
where Einstein summation convention is implied and $\tau^{\gamma}$ are the Pauli matrices. Here, we kept terms only linear in the deformation field and introduced $v$ as the Fermi velocity of the Dirac fermions, which were further gapped by the trivial and topological masses $m$ and $b$. In particular, in the context of the Haldane model, we recognize that $b = 3\sqrt{3}t_2$ for each valley. Having defined a continuum theory for an uncoupled Haldanium bilayer, we further introduce an interlayer coupling under a twisted stacking. In that case, an additional interlayer tunneling term described by
\beq{eq:Ham_tunnel}
    H_{\text{tun}} = \int  \psi^{\dagger}_t T(\Df_{\text{t}} - \Df_{\text{b}}) \psi_b \dd^2\rv \ + \ \text{H.c.},
\eeq
needs to be included, where we expand $T(\Df_{\text{t}} - \Df_{\text{b}}) = \sum_{\vec{G}} T_{\vec{G}} e^{i\vec{G}\cdot (\Df_{\text{t}} - \Df_{\text{b}}) }$, $\vec{G}$ are the reciprocal lattice vectors of the monolayer, and $T_{\vec{G}}$ are the tunneling amplitudes, which depend on the layer separation. Hence, ultimately we arrive at the final expression, introduced in the main text, on combining two terms,
\begin{align}\label{eq:Ham_moire_continuum}
    H_{\text{moir\'e}} = H_{\text{BL}} + H_{\text{tun}}.
\end{align}
The complete continuum Hamiltonian $H_{\text{moir\'e}}$, yields the moir\'e bands $\ket{u^{n}_{\Df,\vec{G}}(\kv)}$ as eigenfunctions, which within the parametrization by the deformation field $\mathcal{D} = (\mathcal{D}_t,\mathcal{D}_l)$, fully encode the local polarization. In particular, the local polarization is heavily-dependent on the stacking-induced deformation field $\mathcal{D}$, influencing the Berry phases in the topological bands obtained from the bilayer Hamiltonian. Namely,
\begin{multline}\label{Eq:dpolarization_cont}
     \P(\Df(\rv)) = \frac{-2ief}{(2\pi)^2} \sum^{\occ}_{n} \int^{\Df(\rv)}_0 \oint_{\text{mBZ}} \braket{ \partial_{\Df} u^{n}_{\Df,\vec{G}}(\kv) | \partial_{\kv} u^{n}_{\Df,\vec{G}}(\kv)} \\ \times d\Df \dd^2 \textbf{k},
\end{multline}
which is expected to change correspondingly across the topological phase transitions controlled by the mass parameters $m$ and $b$, where $b$ combined with the Laplacian act effectively as the further second-neighbor hopping $t_2$.
\bibliography{references.bib}

\end{document}